\documentclass[aip,reprint]{revtex4-1}

\usepackage{graphicx}
\usepackage{dcolumn}
\usepackage{bm}
\usepackage{{subfig}}
\draft 

\newcommand {\bnabla} {\mbox{\boldmath$\nabla$}}

\begin{document}

\title{Field theoretic description of charge-regulation interaction} 

\author{Nata\v sa Ad\v zi\' c}
\affiliation{Department of Theoretical Physics, J. Stefan Institute, 1000 Ljubljana, Slovenia.}
\author{Rudolf Podgornik}
\affiliation{Department of Theoretical Physics, J. Stefan Institute, and
Department of Physics, Faculty of Mathematics and Physics, University of Ljubljana, 1000 Ljubljana, Slovenia.}

\date{\today}

\begin{abstract}
In order to find the exact form of the electrostatic interaction between two proteins with dissociable charge groups in aqueous solution, we have studied a model system composed of two macroscopic surfaces with charge dissociation sites immersed in a counterion-only ionic solution. Field-theoretic representation of the grand canonical partition function is derived and evaluated within the mean-field approximation, giving the Poisson-Boltzmann theory with the Ninham-Parsegian boundary condition. Gaussian fluctuations around the mean-field are then analyzed in the lowest order correction that we calculate {\sl analytically} and {\sl exactly}, using the path integral representation for the partition function of a harmonic oscillator with time-dependent frequency. The first order (one loop) free energy correction gives the interaction free energy that reduces to the zero-frequency van der Waals form in the appropriate limit but in general gives rise to a monopolar fluctuation term due to charge fluctuation at the dissociation sites. Our formulation opens up the possibility to investigate the Kirkwood-Shumaker interaction in more general contexts where their original derivation fails.
\end{abstract}


\maketitle 
\section{Introduction}

Kirkwood and Shumaker more than half a century ago, were the first to realize that there might exist anomalously long-range interactions between proteins in aqueous solutions stemming from thermal charge fluctuations of dissociable charge groups on their surface  \cite{KS1,KS2}. Within the framework of statistical mechanical {\sl perturbation theory} they showed that this interaction is different from the standard van der Waals (vdW) interaction \cite{Pit}, ubiquitous between neutral bodies, primarily because of its extremely long range. The Kirkwood-Shumaker (KS) interaction was shown to scale with a lower inverse power of separation between two proteins then the vdW interaction. Furthermore and contrary to vdW interactions, the KS forces are not universal, but depend on whether and how the protein charge can respond to the local electrostatic potential, a salient property of dissociable charge groups that is usually referred to as {\sl charge regulation} and was first formalized by Ninham and Parsegian \cite{Pars-CR}. 

Charge regulation implies that the effective charge on a macroion, e.g. protein surface, responds to the local solution conditions, such as local $pH$, local electrostatic potential, salt concentration, dielectric constant variation and most importantly the presence of other vicinal charged groups \cite{Lund}. Although charge regulation is an old concept, modern theories of electrostatic interaction between macroions immersed in Coulomb fluids \cite{RudiandCo} mostly deal with constant surface charge of a macroion, bypassing the complications introduced by charge regulation \cite{Chan,Boon,Netz-CR}. Constant charge is of course a very stringent approximation and holds only in a very restricted part of the parameter space. In general, however, the charge of a macroion surface with dissociable groups always depends strongly on the acid-base equilibrium that defines the fraction of acidic (basic) groups that are dissociated \cite{BoJ}, and it is necessary to incorporate this property consistently into a theoretical formulation. Our goal in this work is thus to find a theoretical description which would take into account charge regulation of dissociable surface groups and would allow to generalize the original Ninham-Parsegian derivation to include the contribution of fluctuations around the mean field, as well as to pave the way towards other approximations that go beyond the simple mean field {\sl Ansatz}.

We will first show what is the correct free energy that corresponds to the Ninham-Parsegian mean-field charge regulation theory \cite{Pars-CR}. It will furthermore become clear as we proceed that the KS interactions in fact correspond to Gaussian monopolar charge fluctuations around the Ninham-Parsegian state, different from the dipolar fluctuations at the origin of the standard vdW forces. We will derive explicitly an exact expression for the one-loop correction of the free energy in the case of a counterion-only system in a planar parallel slab geometry. The theory presented here, while being explicitly formulated only for a restricted model, allows for many generalizations of the monopolar fluctuation interactions that can be derived within the same formalism. We will nevertheless not pursue these generalizations here and relegate further developments to subsequent publications.

The plan of the paper is as follows: in Section \ref{sec:model} we start from the simplest model that retains the salient features of charge regulation, composed of two planar parallel macromolecular surfaces with surface distributed charge dissociation sites, immersed in a Coulomb fluid composed of counterions only. We  base our analysis on a field theoretic description of the system's partition function, whose Hamiltonian is generalized to include a surface term which describes properly the charge regulation and consequently the local charge fluctuation at the macromolecular surfaces. This model is introduced in Section \ref{subsec:ftm} together with its full free energy and is shown to coincide with the Ninham-Parsegian {\sl Ansatz} for the charge dissociation equilibrium on the mean-field  level and to reduce to the Poisson-Bolzmann (PB) equation with the charge-regulation boundary condition in Section \ref{subsec:mf}. In Section \ref{subsec:soc} we address the Gaussian fluctuations around this mean-field solution with its charge-regulation boundary condition that can in fact be solved {\sl exactly}  and {\sl analytically}. The exact one-loop free energy correction is obtained by using the path-integral approach for a harmonic oscillator with time-depended frequency with all the relevant technical details relegated to Appendix \ref{sec:ap1}. Finally in Section \ref{subsec:nmr} we present numerical results and comment upon its relevance for the KS interaction in the Conclusions Section \ref{sec:con}.

 \begin{figure}[!t]
\centering{\includegraphics[width=0.45\textwidth]{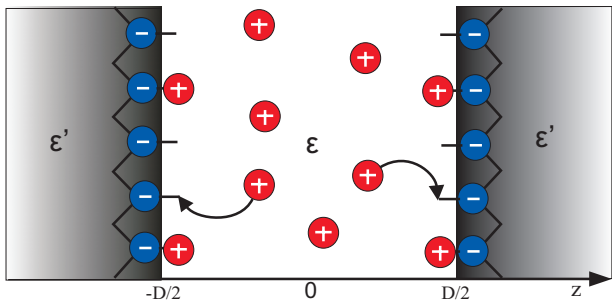}}
\caption{Schematic representation of two charged planar surfaces at a separation $D$ with charge dissociation sites distributed uniformly along the surfaces and with counterions between the surfaces. The counterions originate from the charge dissociation of the dissociable groups ($AC$) through the reaction $\rm AC \leftrightarrow A^{-} + C^{+}$.}
\label{fig:fig0}
\end{figure}

\section{The model}\label{sec:model}

We consider two flat parallel plates, located at $z = \pm D/2$ and immersed into an aqueous solvent, that carry dissociable charge groups of the type $\rm AC \leftrightarrow A^{-} + C^{+}$, where the counter ion $C$ is released into the aqueous solution,   Fig. \ref{fig:fig0}. We do not specify the identity of the released counterion but assume it is the only mobile species in the considered model. Furthermore we assume a grand canonical ensemble for the counter ions, specified by a fixed value of the activity. The number of the counter ions in the solution is thus not fixed but depends on the dissociation state of the surfaces. While in standard formulations of the counter ion-only Coulomb fluids with fixed boundary charge the grand canonical formulation is just a step towards the final canonical ensemble, corresponding to a fixed number of charges, in our case this is not fixed and the grand canonical description is natural. 

We need to note that in the Ninham-Parsegian model the released counter ion is a proton and the aqueous solution contains a salt mixture at a specified ionic strength for both monovalent and divalent complements \cite{Pars-CR}. While this model can be formalized in the same way as our simplified model, we first solve the simplified case in order to derived the proper level of description as well as to investigate the salient features of fluctuations in a case, where they can be treated exactly. 

In order to describe the surface charge dissociation we introduce a lattice gas model with its own surface free energy contribution. This surface part of the free energy stems from the charge dissociation equilibrium and describes the (free) energy penalty for a finite surface charge density. We show furthermore that on the mean-field level our formulation yields {\sl exactly} the same result as the Ninham-Parsegian charge regulation {\sl Ansatz}, which is not explicitly based on any surface free energy.  The equilibrium distribution of the counter ions is then obtained from the saddle-point equation, that corresponds to the minimum of the complete, i.e. volume plus surface, free energy. The dielectric constant in the region between the walls is taken as $\ \epsilon $, while outside that region it is assumed to be in general different and equal to $\ \epsilon ' $.\\

\section{Field theoretic description of the model}\label{subsec:ftm}

For describing this model system of interacting particles it is advantageous to use the field-theoretic formalism to derive the partition function. The configurational part of the Hamiltonian of an auxiliary system of $N$ counter ions, with a fixed surface charge density $\sigma _0$ on the bounding surfaces, can be written as
\begin{equation}
H=\frac{1}{2}\sum_{i\neq j} u(\vec{r}_{i},\vec{r}_{j})e_{i}e_{j}+\sum_{i=1}^{N}\oint u(\vec{r},\vec{r}_{i})\sigma _0 d^{2}\vec{r} ,
\end{equation}
where $\oint $ implies an integration over all the charged bounding surfaces and $\ u(\vec{r},\vec{r}_{i})$ is the electrostatic interaction kernel, i.e. Green's function of the Coulomb potential, which satisfies the relation
\begin{equation}
\bnabla^2 u(\vec{r},\vec{r}_{i})=-\frac{\delta (\vec{r}-\vec{r}_{i})}{\epsilon \epsilon _{0}}.
\end{equation}
The canonical configurational partition function of the system can then be represented by an integral over all positions of the counterions
\begin{equation}
Q_{N}=\int d\vec{r}_{1}...d\vec{r}_{N}e^{-\beta H}.
\end{equation}
After applying the  Hubbard-Stratonovich transformation, one can obtain the grand canonical partition function as a functional integral over the fluctuating electrostatic potential $\ \varphi (\vec{r})$
\begin{equation}
{\cal Z}=\int\mathcal{D}[\varphi (\vec{r})] ~e^{-{\cal S}[\varphi (\vec{r})]}, \label{eq:gcpf}
\end{equation}
with the field-action of the form:
\begin{eqnarray}
{\cal S}[\varphi (\vec{r})] &=& \frac{1}{2}\beta \epsilon \epsilon _{0}\int d^{3}\vec{r}~\vert\nabla \varphi (\vec{r})\vert^{2}+ \tilde{\lambda} \int d^{3}\vec{r}~e^{i\beta e \varphi (\vec{r})} + \nonumber\\
&+& i\beta \oint d^{2}\vec{r}~\sigma _0 \varphi (\vec{r}),
\label{eq:S}
\end{eqnarray}
Here $\tilde{\lambda}$ is the absolute activity that will be obtained self-consistently. The above field-action is universal in terms of the non-linear volume interaction term, the second term in the above equation, that corresponds exacty to the van't-Hoff ideal osmotic pressure of the counter ions.  This is a well-known result \cite{funint}, which on the weak coupling mean-field level, using substitution $\ \varphi \to i\phi _{MF}$, gives the PB equation with fixed charged density boundary condition $\ \vec{n}\cdot\vec{\nabla }\phi _{MF}=\sigma _0$ \cite{RudiandCo}. 

We now generalize this free energy {\sl Ansatz} so that it will contain also a surface part, not necessarily linear in the surface fluctuating potential, by assuming that the surface free energy in Eq. \ref{eq:S} can be modified as
 \begin{equation}
i\oint \sigma _0 \varphi (\vec{r})d^2\vec{r} \longrightarrow  \oint f(\varphi (\vec{r}))d^2\vec{r},
\label{gbefhwjk}
\end{equation}
where $\ f(\varphi (\vec{r}))$ is a general non-linear function of the local potential. The exact form of this surface free energy  is not universal and depends on the model of the surface-ion interaction \cite{Tomer}. Here, we will delimit ourselves to a surface lattice gas model, which was introduced in a different context by Fleck and Netz \cite{FleckNetz},  and derive the corresponding free energy, as well as show that the same model in fact corresponds exactly to the Ninham-Parsegian charge regulation theory \cite{NP}.  The surface lattice gas model of dissociable charged groups gives \cite{lattice,FleckNetz}
\begin{equation}
f(\varphi({\bf r})) = ~  i \sigma_0 \varphi({\bf r}) - k_BT~\frac{\mid{\sigma_0}\mid}{e_0} \ln{\left( 1 + e^{ \beta \mu _S + i\beta e_0 \varphi({\bf r})} \right)}, \label{eq:sfsf}
\end{equation}
where $\mu _S $ is the free energy of dissociation. In the argument of the logarithm function one can recognize the partition function for a system with uncharged ground state and a charged state with an effective energy $\ \beta \mu _S + i\beta e\varphi (\vec{r})$. It is possible to generalize this model with other surface free energies \cite{Andelman,Dean} that can capture other details of the surface-ion interaction. Furthermore, in the limit of $\beta \mu _S \longrightarrow \infty$, the sites are completely undissociated,  the bounding surfaces are uncharged and there is no contribution to the surface free energy. In the opposite limit, $\beta \mu _S \longrightarrow -\infty$, the bounding surfaces are completely dissociated and we are back to the fixed surface charge $f(\varphi({\bf r})) = ~  i \sigma_0 \varphi({\bf r})$. 

The complete field action of the model at hand thus assumes the form
\begin{eqnarray}
& & {\cal S}[\varphi (\vec{r})] = \frac{1}{2}\beta \epsilon \epsilon _{0}\int d^{3}\vec{r}~\vert\nabla \varphi (\vec{r})\vert^{2} + \tilde{\lambda} \int d^{3}\vec{r}~e^{i\beta e \varphi (\vec{r})} + \nonumber\\ 
 & & i \beta \oint d^2\vec{r} \sigma_0 \varphi({\bf r}) - \oint d^2\vec{r} \frac{\mid{\sigma_0}\mid}{e_0} \ln{\left( 1 + e^{ - \beta \mu _S + i\beta e_0 \varphi({\bf r})} \right)}.\nonumber\\
~
\label{eq:S1}
\end{eqnarray}
While the volume part presents an exact field-theoretic representation of the counter ion partition function, the surface part pertains to a specific model of the interaction between the mobile charges and the bounding surfaces.

\section{Mean field approximation}\label{subsec:mf}

The functional integral Eq. \ref{eq:gcpf}, with the field-action functional $\ {\cal S}[\varphi (\vec{r})] $ decomposed as 
\begin{equation}
  {\cal{S}}[\varphi (\vec{r})] =\int_V f_V(\varphi (\vec{r})) ~d^3r+ \oint_S f_S(\varphi (\vec{r})) ~d^2r, \label{bcfghjsek}
\end{equation}
can not be evaluated exactly, since it is in general not Gaussian. One thus has to take recourse to various approximations of which the mean-field approximation, being equivalent to the saddle-point approximation, is the most straightforward one. 

The mean-field potential $\phi_{MF} (\vec{r})$ of the field-action Eq. \ref{bcfghjsek} is defined as a solution of the saddle-point equation corresponding to $\delta{{\cal{S}}[\varphi (\vec{r})]} = 0$ at $\varphi (\vec{r}) = i \phi_{MF} (\vec{r})$ where $\phi_{MF} (\vec{r})$ is thus a solution of
\begin{eqnarray}
&&\bnabla \left( \frac{\partial f_V(\phi_{MF} (\vec{r}))}{\partial \bnabla \phi_{MF} (\vec{r})}\right) - \frac{\partial f_V(\phi_{MF} (\vec{r}))}{\partial \phi_{MF} (\vec{r})}= 0 
\end{eqnarray}
{\rm and}
\begin{eqnarray}
 &&- \beta \epsilon\epsilon_0 \frac{\partial \phi_{MF}({\vec r} )}{\partial {\vec n}} = \frac{\partial f_S(\phi_{MF}({\vec r} ))}{\phi_{MF}({\vec r})} = \sigma(\phi_{MF}({\vec r})),
 ~\label{eq:sppb}
 \end{eqnarray}
where $\ {\vec n}$ is the normal vector to the bounding surface(s), and $\ \sigma (\phi_{MF}({\vec r} ))$ is the effective surface charge at the bounding surface(s). {\sl In extenso} the first equation is exactly the standard PB equation for the counterion only system
\begin{equation}
\nabla ^{2}\phi_{MF}(\vec{r}) =-\frac{\tilde{\lambda }e }{\epsilon \epsilon _{0}}~e^{-\beta e \phi_{MF}(\vec{r})} ,\label{eq:onecomponentPB}
\end{equation}
while the second saddle-point equation with $f(\varphi({\vec r}))$ from Eq. \ref{eq:sfsf} reduces to the boundary condition
\begin{eqnarray}
& & - \beta \epsilon\epsilon_0 \frac{\partial \phi_{MF}({\vec r} )}{\partial {\vec n}} 
=  -\frac{\sigma_0}{2} \left( 1 + \tanh{{\textstyle\frac12} \left( -\beta \mu_S + \beta e_0 \phi_{MF}\right)}\right).\nonumber\\
~
\label{biwy}
\end{eqnarray}
Obviously the above surface charge density can span the interval $\ [-\sigma _0, 0]$.

Assuming that $\ \beta \mu _S= - \ln 10(pH-pK)$, with $\  \rm pK = - \log{K} $ and $K$ being the dissociation equilibrium constant while  $\  \rm pH = - \log{[H^+]} $ with $[H^+]$ the concentration of the protons in the bath, the above boundary condition coincides exactly with the charge regulation boundary condition of the  Ninham-Parsegian site-dissociation  model \cite{NP}. Should there be more then one type of dissociable groups the proper generalization was introduced in Ref. \cite{vonGru}.

For the planar geometry the mean-field solution of Eq. \ref{eq:onecomponentPB} depends only on the $z$ coordinate and has the form
\begin{equation} 
\phi_{MF}(z)=\frac{1}{\beta e}\ln{[\cos^2{(\alpha z)}]},
\end{equation}
where $\ \alpha $ can be determined from the boundary condition Eq. \ref{biwy} as
\begin{equation}
(1+b)\alpha \tan{(\alpha D/2)}+b\alpha \tan^3{(\alpha D/2)}=\frac{1}{\mu },\label{eq:bcasass3}
\end{equation}
with $\ b$ being related to the dissociation free energy as $\ln{b}= \beta \mu _S$. Here $\ \mu $ is the Gouy-Chapman length, which represents the characteristic distance at which a counterion interacts with a macromolecular flat surface, of surface charge $\ \sigma _0$, with an energy $\ k_{B}T$ and is defined as $\ \mu =2\epsilon _0\epsilon /e\beta \sigma_0 $. 

\section{Second order (Gaussian) correction}\label{subsec:soc}

After solving the mean-field equations, one proceeds to analyze the fluctuations around the mean-field potential by evaluating the partition function  Eq. \ref{eq:gcpf} for the field-action functional $\ {\cal S}[\phi (\vec{r}) = \phi_{MF}(\vec{r}) + \delta \phi (\vec{r})] $. To the lowest Gaussian order in the field fluctuations  $\delta \phi (\vec{r})$ the field-action can be expanded
\begin{equation}
{\cal S}[\phi (\vec{r})] = {\cal S}[ \phi_{MF}(\vec{r}) + \delta \phi (\vec{r})] = S_{MF}[\phi_{MF}] + {\cal S}_2[\delta\phi (\vec{r})]
\end{equation}
where 
\begin{eqnarray}\label{eq:dec}
{\cal S}_2[\delta\phi (\vec{r})] &&=\frac{1}{2}\int \int \frac{\delta ^{2}S}{\delta \phi (\vec{r})\delta \phi (\vec{r})}\vert_{MF}~\delta \phi (\vec{r})\delta \phi (\vec{r'})d^{3}\vec{r}d^{3}\vec{r'}+\nonumber\\
&&+ {\textstyle\frac{1}{2}}\oint ~{\cal C}_{S}(\phi (\vec{r'}))\vert_{MF} \delta \phi (\vec{r'})^2 d^{2}\vec{r},
 \end{eqnarray}
and obviously decomposes into a volume and surface term just like the complete field action.  Above we introduced the Hessian of the volume part of the field-action as
\begin{equation}
 {\textstyle\frac12}  \frac{\delta ^{2}S}{\delta \phi (\vec{r})\delta \phi (\vec{r})}\vert_{MF} = {\textstyle\frac12} \beta\left( u^{-1}(\vec{r},\vec{r}') -  \frac{\beta \tilde{\lambda} }{\cos^2{\alpha z}}~\delta^3(\vec{r} - \vec{r}')  \right);
\end{equation}
while ${\cal C}_S$ is the surface capacitance due to the nonlinear coupling of surface charge and surface electrostatic potential
\begin{equation}
{\cal C}_S({\vec r}) = \frac{\partial^2 f(\phi _{MF}({\vec r}))}{\partial (\beta e \phi _{MF}({\vec r}))^2} = \frac{\partial \sigma}{\partial (\beta e \phi _{MF}({\vec r}))}.
\end{equation}
We will show later on that in the original theory of  KS interactions it is this surface capacitance that quantifies the thermal charge fluctuations \cite{Lund}.\\

The decomposition of the field action Eq. \ref{eq:dec} induces a decomposition of the partition function into a product of the saddle-point partition function and its first order correction, so that finally
\begin{eqnarray}\label{eq:z3}
{\cal Z} &=&e^{-\frac{1}{2}\ln{[\det{\beta u(\vec{r},\vec{r'})}]}} \times e^{{\cal S}[\phi_{MF} (\vec{r})]} \times \nonumber\\
& & \times\int\mathcal{D}[\delta\phi (\vec{r})] ~e^{{\cal S}_2[\delta\phi (\vec{r})]} = {\cal Z}_{MF} \times {\cal Z}_2.
\end{eqnarray} 
The last term is due to Gaussian fluctuation around the saddle point and thus corresponds to the one-loop correction in the free energy. 

In order to proceed we first introduce the appropriate field Green's function
\begin{eqnarray}
&&{\cal G}\Big({\delta \phi_1}(\vec{r}), {\delta \phi_2}(\vec{r})\Big) =\nonumber\\
&& \int_{{\delta \phi_1}}^{{\delta \phi_2}}\mathcal{D}[\delta\phi (\vec{r})] ~e^{\frac{1}{2}\int\!\!\!\int \frac{\delta ^{2}S}{\delta \phi (\vec{r})\delta \phi (\vec{r})}\vert_{MF}~\delta \phi (\vec{r})\delta \phi (\vec{r'})d^{3}\vec{r}d^{3}\vec{r'}}
\end{eqnarray}
that describes the field, or better the propagation of Gaussian electrostatic potential fluctuations and will allow us to formally separate the bulk and the surface terms in the calculation of the one-loop partition function.

Since  the kernel $u^{-1}(\vec{r},\vec{r}')$ is isotropic in the transverse directions $\rho = (x,y)$, one can introduce the Fourier-Bessel transform of the fluctuating potential as
\begin{equation} 
\delta \phi (\vec{r}) = \delta \phi (\rho ,z) = \int_{0}^{\infty }dQ J_{0}(Q\rho ) \delta \phi(Q,z), 
\end{equation}
where $\delta \phi(Q,z)$ depends only on the magnitude of the 2D transverse wave vector, $Q = \vert {\bf Q}\vert$.  With this notation  the complete Green's function can be presented as the product 
\begin{equation}
{\cal G}\Big({\delta \phi_1}(\vec{r}), {\delta \phi_2}(\vec{r})\Big) = \Pi_Q {\cal G}_Q\Big({\delta \phi}(Q,z_1), {\delta \phi}(Q,z_2)\Big),
\end{equation}
where $\  {\cal G}_Q\Big({\delta \phi}(Q,z_1), {\delta \phi}(Q,z_2)\Big)$ can be furthermore derived in the form
\begin{widetext}
\begin{eqnarray}\label{eq:pathkl}
&& {\cal G}_Q\Big({\delta \phi}(Q,z_1), {\delta \phi}(Q,z_2)\Big)\!\!=\!\!\int_{{\delta \phi(Q,z_1)}}^{{\delta \phi(Q,z_2)}}\!\!\!\!\!\!\!\!\!\!\!\!\mathcal{D}[\delta\phi (Q,z)] \exp{\Big[-{\textstyle\frac{1}{2}} \beta \epsilon\epsilon_0} { \int_{z= z_1}^{z= z_2}\!\!\!\!\!\!\!\!\!\!\!dz ~\Bigg( \left( \frac{d ~\delta \phi }{dz}\right)^2 \!\!\!\!-\!\! \left( Q^2 + \frac{2\alpha ^2}{\cos^2{(\alpha z)}}\right)\!\!\delta \phi^2 \Bigg)\Big]}. \label{neklr}
\end{eqnarray}
\end{widetext}
Obviously this is nothing but the Feynman propagator of a harmonic oscillator with time-depended frequency, where the $\ z$ coordinate plays the role of "time"  \cite{Feynman}, and the Wick's rotation makes the action real instead of imaginary as in quantum mechanics.  The general method of solving this type of functional integrals was described by Khandekar and Lawande \cite{KL} and was adapted to this particular case as described in detail in the Appendix \ref{sec:ap1}. 

The partition function, or specifically the part stemming from Gaussian fluctuations, Eq. \ref{eq:z3}, around the mean-field can now be cast into the following form
\begin{widetext}
\begin{eqnarray}
&&{\cal Z}_2(D) = \Pi_Q~\int\mathcal{D}[\delta\phi_1 (\vec{r}) \delta\phi_2 (\vec{r})]  ~  \nonumber\\
&& \tilde{\cal G}_Q\Big(0, {\delta \phi_1}(\vec{r}) \Big) \times e^{- \frac{1}{2}\int_{S_1}\!\!d^2r {\cal C}_{S_1}(\phi_{MF} ) {\delta \phi_1}^2(\vec{r})}\times {\cal G}_Q\Big({\delta \phi_1}(\vec{r}), {\delta \phi_2}(\vec{r})\Big) \times e^{- \frac{1}{2} \int_{S_2}\!\!d^2r {\cal C}_{S_2}(\phi_{MF} ) {\delta\phi_2}^2(\vec{r})}  \times  \tilde{\cal G}_Q\Big({\delta \phi_2}(\vec{r}), 0 \Big), \nonumber\\
~
\label{negiorusnhi11}
\end{eqnarray}
\end{widetext}
where $\tilde{\cal G}_Q$ stands for the Green's function Eq. \ref{neklr} but with $\alpha = 0$, as there are no counterions behind the two bounding surfaces. The exact form Eq. \ref{eq:gfprop} thus still remains valid but evaluated explicitily for vanishing $\alpha$. Of course in that case the functional integral can be evaluated directly in a trivial fashion. In addition, one needs to take the dielectric constant as $\epsilon'$ for $\tilde{\cal G}_Q\Big(0, {\delta \phi_1}(\vec{r}) \Big) $ and $\tilde{\cal G}_Q\Big({\delta \phi_2}(\vec{r}), 0 \Big)$, but as $\epsilon$ for ${\cal G}_Q\Big({\delta \phi_1}(\vec{r}), {\delta \phi_2}(\vec{r})\Big)$ in the definition Eq. \ref{eq:pathkl}. 

One could see the above formula as describing fluctuations behind the surface at $z = z_1$, described by $\tilde{\cal G}_Q\Big(\epsilon'; 0, {\delta \phi_1}(\vec{r}); \infty \Big)$, fluctuations behind the surface at $z = z_2$, described by $\tilde{\cal G}_Q\Big(\epsilon'; {\delta \phi_2}(\vec{r}), 0); \infty \Big)$, fluctuations in the space between the two surfaces for $z_1 < z < z_2$, described in their turn by ${\cal G}_Q\Big(\epsilon; {\delta \phi_1}(\vec{r}), {\delta \phi_2}(\vec{r}); D\Big)$, and finally all of them coupled through the surface capacitance and the surface potential fluctuations at the two surfaces at $z = z_1$ and $z = z_2$ corresponding to the two exponential terms. 

After integration over the boundary electrostatic potential fluctuations the final exact form of the partition function can be written as 
\begin{widetext}
\begin{eqnarray}\nonumber
&&{\cal Z}_2(D) =  \Pi_Q~\sqrt{\frac{2e^{-DQ}Q(\alpha ^2+Q^2)}{2\pi (\alpha \tan{[\alpha D/2]}+Q)^2- (\alpha \tan{[\alpha D/2]}-Q)^2e^{-2DQ})}} \times\nonumber\\
&& \sqrt{\frac{1}{{\cal C}_{S_1}{\cal C}_{S_2} + \beta \epsilon'\epsilon_0 ({\cal C}_{S_1} + {\cal C}_{S_2}) Q+(\beta \epsilon\epsilon_0)^2 N^2 + (\beta \epsilon'\epsilon_0)^2 Q^2 + (\beta \epsilon\epsilon_0) ({\cal C}_{S_1} + {\cal C}_{S_2}+ 2\beta \epsilon'\epsilon_0 Q) M}};
\end{eqnarray}
with the functions $\ M$ and $\ N$ defined as
\begin{eqnarray}
&&M=\frac{Q(\alpha \tan{[\frac{\alpha D}{2}]}+Q)^2+(\alpha ^2+\alpha ^2\tan^2{[\frac{\alpha D}{2}]})(\alpha \tan{[\frac{\alpha D}{2}]}+Q))}{(\alpha \tan{[\frac{\alpha D}{2}]}+Q)^2- (\alpha \tan{[\frac{\alpha D}{2}]}-Q)^2e^{-2DQ}}-\nonumber\\
&&-\frac{Q(\alpha \tan{[\frac{\alpha D}{2}]}-Q)^2-(\alpha ^2+\alpha ^2\tan^2{[\frac{\alpha D}{2}]})(\alpha \tan{[\frac{\alpha D}{2}]}-Q))e^{-2DQ}}{(\alpha \tan{[\frac{\alpha D}{2}]}+Q)^2- (\alpha \tan{[\frac{\alpha D}{2}]}-Q)^2e^{-2DQ}};\nonumber\\
&&N^2=M^2-\frac{4e^{-2DQ}Q^2(\alpha ^2+Q^2)^2}{\Big[(\alpha \tan{[\frac{\alpha D}{2}]}+Q)^2- (\alpha \tan{[\frac{\alpha D}{2}]}-Q)^2e^{-2DQ}) \Big]^2}.\nonumber\\
~
\end{eqnarray}
\end{widetext}
We have thereby derived the explicit and exact forms of the partition function in the form of a mean-field term and the one-loop or Gaussian fluctuation correction that has not been calculated before. 

What remains now is the evaluation of the corresponding free energy and specifically the part of this free energy that depends on the separation between the bounding surfaces, i.e. the interaction free energy.

\section{Second order correction - interaction free energy}\label{subsec:soc}

Knowing the partition function for Gaussian fluctuations around the mean-field, one can straightforwardly calculate the second-order  or the one-loop correction to the free energy as
\begin{equation}
 \frac{{\cal F}_2(D)}{S}=-k_BT\ln {\frac{{\cal Z}_2(D)}{{\cal Z}_2(D\to \infty)}},
 \end{equation}
where we subtracted the free energy corresponding to infinite separation that contains the bulk free energy as well as the surface self-energies.  

Assuming furthermore that the surfaces have identical properties, i.e., $\ {\cal C}_{S_1}={\cal C}_{S_2}={\cal C}_{S}$ we get the one-loop correction as:
\begin{widetext}
\begin{eqnarray}\label{eq:tr3dksl}
&&\frac{{\cal F}_2(D)}{S}=\frac{k_BT}{4\pi} \int_0^\infty QdQ~ \log{\Big[\frac{1}{(\alpha ^2+Q^2)} \Delta ^{2}_{11}(Q)\Big]}+\frac{k_BT}{4\pi} \int_0^\infty QdQ~ \log{\Big( 1 - \Delta ^2_{12}(Q) ~e^{- 2 QD }\Big)}:\nonumber\\
~
\end{eqnarray}
where we defined the following quantities
\begin{equation}
\Delta _{11}(Q)=\frac{{\cal C}_{S}(\alpha \tan{[\alpha D/2]}+Q)+\beta \epsilon_0[\epsilon 'Q(\alpha \tan{[\alpha D/2]}+Q)+\epsilon \{Q(\alpha \tan{[\alpha D/2]}+Q)+(\alpha ^2+\alpha ^2\tan^2{[\alpha D/2]})\}]}{{\cal C}_{S}+\beta \epsilon_0Q(\epsilon '+\epsilon )};\nonumber\\
~
\end{equation}
\begin{equation}
\Delta _{12}(Q)=\frac{{\cal C}_{S}(\alpha \tan{[\alpha D/2]}-Q)+\beta \epsilon_0[\epsilon 'Q(\alpha \tan{[\alpha D/2]}-Q)-\epsilon \{Q(\alpha \tan{[\alpha D/2]}-Q)-(\alpha ^2+\alpha ^2\tan^2{[\alpha D/2]})\}]}{{\cal C}_{S}(\alpha \tan{[\alpha D/2]}+Q)+\beta \epsilon_0[\epsilon 'Q(\alpha \tan{[\alpha D/2]}+Q)+\epsilon \{Q(\alpha \tan{[\alpha D/2]}+Q)+(\alpha ^2+\alpha ^2\tan^2{[\alpha D/2]})\}]}.\nonumber\\
~
\end{equation}
\end{widetext}
The second order correction free energy Eq. \ref{eq:tr3dksl} consists of two integrals. The first one corresponds to that part of the self-energy of the two bounding surfaces that depends on the inter surface separation, while the second integral represents a generalization of the zero-frequency (classical) vdW-Lifshitz term \cite{funint2}. In fact it can be easily seen that in the limit of no mobile ions between the surfaces, corresponding to $\ \alpha =0$, it reduces exactly to the zero-frequency vdW term with 
\begin{equation}
\Delta _{12}^2(Q)=\Big(\frac{\epsilon '-\epsilon}{\epsilon '+\epsilon }\Big)^2,
\end{equation}
while the first term vanishes. With mobile ions present, the second order correction is however very different from this limit.  In the limit of fixed surface charge ($\ {\cal C}_{S_1}={\cal C}_{S_2}=0$) and no dielectric  discontinuity ($\ \epsilon '=\epsilon $), the integral reduces to the known result \cite{Matej}:
\begin{eqnarray}\label{eq:trans3}
&&\frac{{\cal F}_2(D)}{S}=\frac{k_BT}{4\pi} \int_0^\infty \tilde{Q}d\tilde{Q}~ \log{\Big[\frac{1}{(\tilde{\alpha }^2+\tilde{Q}^2)}}\times\nonumber\\
&&\times\Big(\frac{2\tilde{Q}+2\tilde{Q}^2+\tilde{\alpha }^2 +1}{2\tilde{Q}}\Big)^2\Big]\nonumber\\
&&+\frac{k_BT}{4\pi} \int_0^\infty \tilde{Q}d\tilde{Q}~ \log{\Big( 1 - \Big(\frac{1+\tilde{\alpha }^2}{2\tilde{Q}+2\tilde{Q}^2+\tilde{\alpha }^2 +1}\Big)^2e^{-2\tilde{D}\tilde{Q}}\Big)},\nonumber\\
~
\end{eqnarray}
leading to the attractive pressure which scales as $\log{\tilde{D}} \times \tilde{D}^{-3}$ in a system composed of mobile counterions and fixed surface charge. At the end we also consider a formal limit of the free energy corresponding to no dielectric discontinuity $\ \epsilon'=\epsilon$, as well as no mobile ions $\ \alpha \to 0$, but nevertheless assuming a non-vanishing surface capacitance $\ {\cal{C}}$. While this limit is not meaningful in our model, we will nevertheless use it to show how the KS result \cite{KS1,KS2}, which is based on a linear response formalism and considers no coupling between the mean-field solution and the corresponding values of the capacitances, is obtained from our conceptual framework. 

The KS limit could be obtained more directly if instead of a counterion-only case dealt with here, we would consider a uni-univalent salt as indeed was considered by Kirkwood and Shumaker in their derivation of the long range interaction between protein molecules with dissociable surface groups \cite{KS1,KS2}. Nevertheless, for $\ \alpha \to 0$ our general result reduces to 
\begin{equation}
\Delta _{12}^2(Q)=\big(\frac{{\cal{C}}}{{\cal{C}}+2\beta \epsilon \epsilon _0Q}\Big)^2,
\end{equation}
which in its turn, to the lowest order in the surface capacitance leads to the disjoining pressure 
\begin{equation}
p=-\frac{\partial }{\partial D}\Big(\frac{{\cal F}_2(D)}{S}\Big) \sim {\cal{C}}^2 D^{-1}.
\label{fngewh}
\end{equation} 
As it depends quadratically on the surface capacitance, this interaction presents the contribution of monopolar fluctuations in the surface charge to the free energy. This can be easily confirmed by evaluating the free energy of two fluctuating charge distributions in the Gaussian approximation explicitly. Let us now show that interaction pressure Eq. \ref{fngewh} corresponds exactly to the KS interaction between two planar surfaces.

\begin{figure*}[!t]
\subfloat[$\ $]{\includegraphics[width=0.5\textwidth]{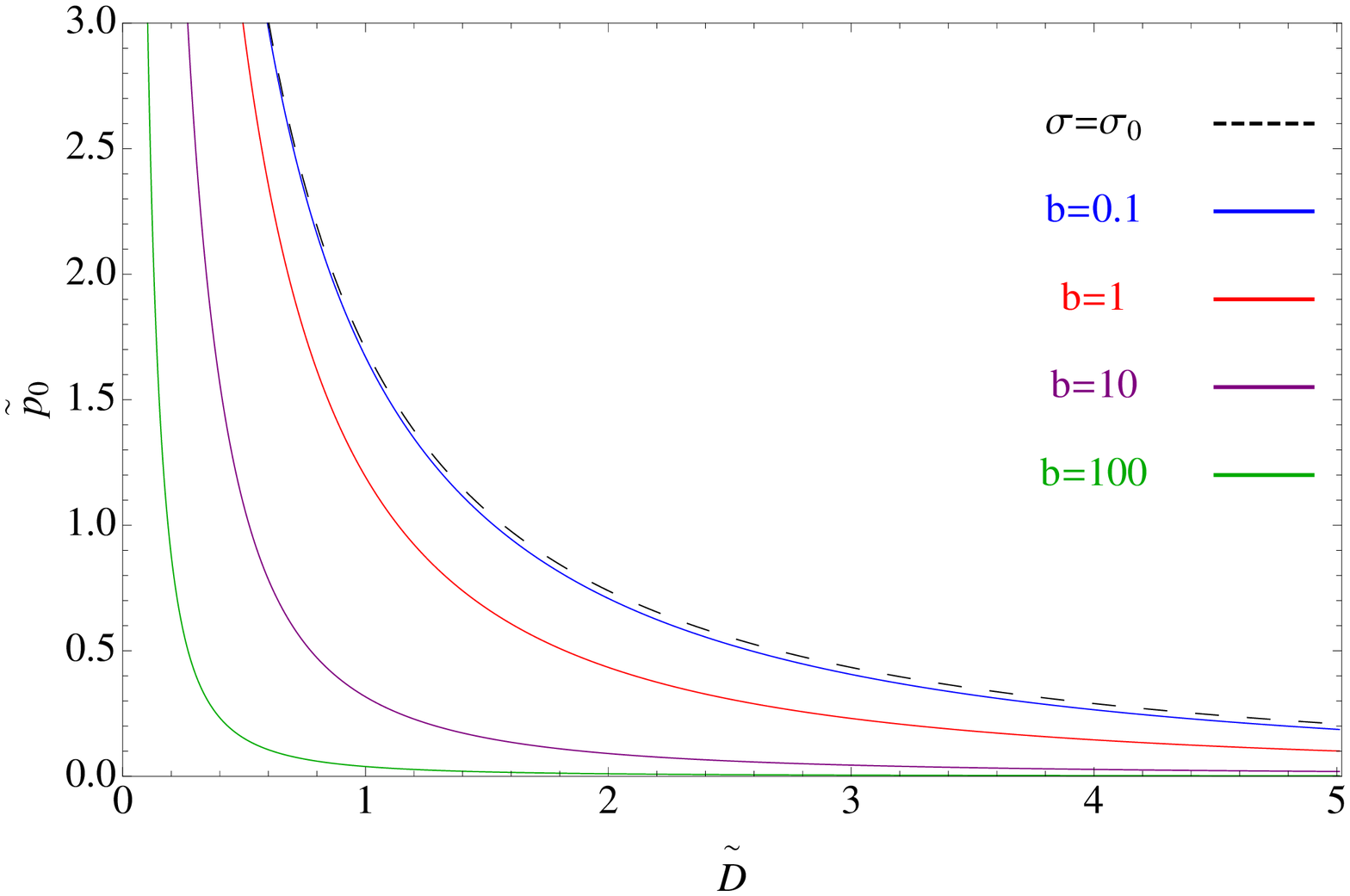}}\subfloat[$\ $]{\includegraphics[width=0.5\textwidth]{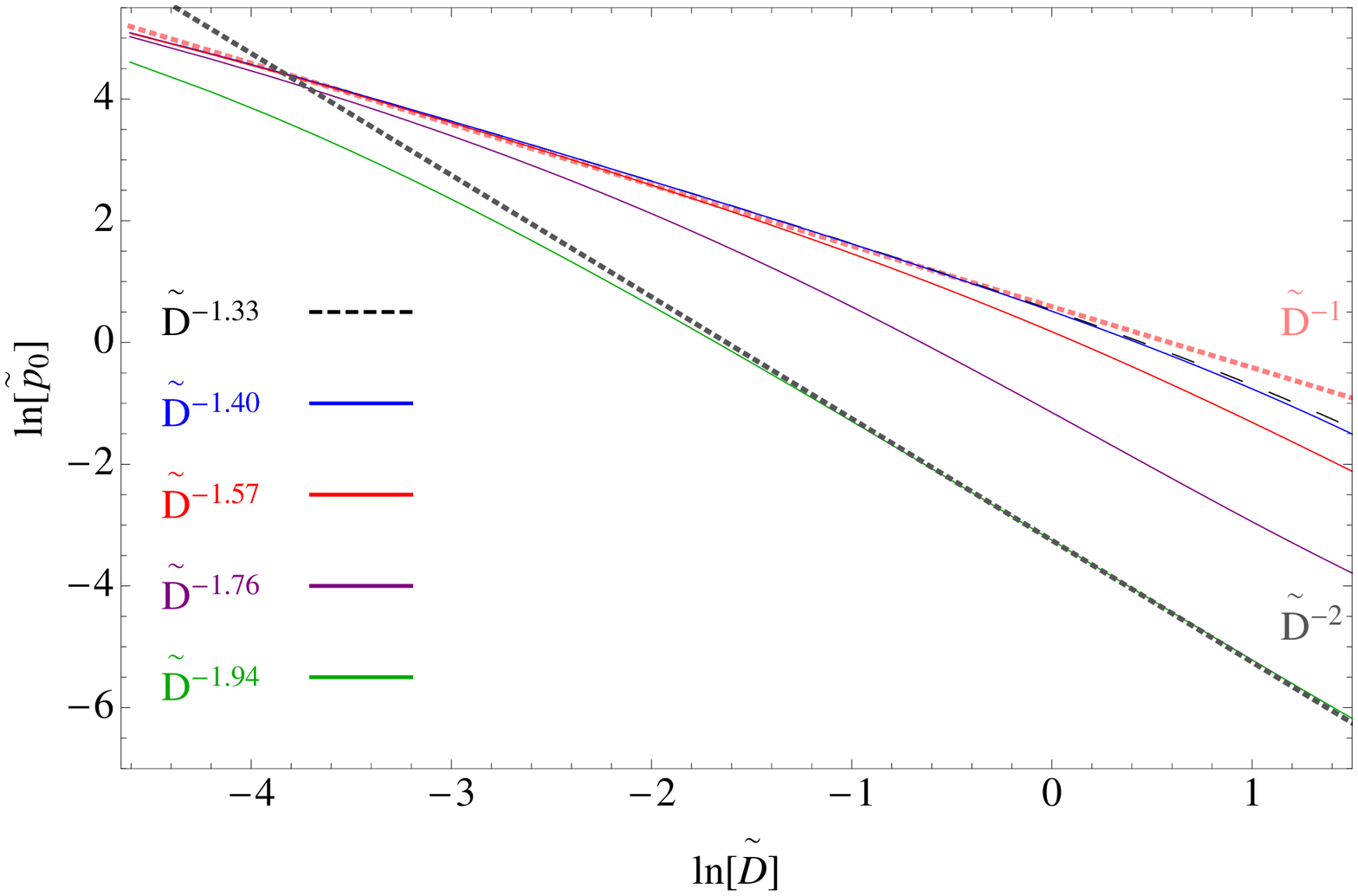}}
\caption{(a) Rescaled mean-field disjoining pressure plotted as a function of rescaled surfaces separation for different values of parameter $\ b$. The curve $\sigma = \sigma_0$ corresponds to $b = 0$.  (b) Rescaled mean-field pressure from (a) plotted in a log-log plot. The two dotted lines represent the scalings ${\tilde D}^{-1}$ and ${\tilde D}^{-2}$ introduced solely to guide the eye. Obviously the scaling ${\tilde D}^{-1}$ for mean-field pressure sets in for small and ${\tilde D}^{-2}$ for large values of the dimensionless separation. }
\label{fig:fig1}
\end{figure*}

In fact, the disjoining pressure Eq. \ref{fngewh} starts to become more familiar when we realize that a Hamaker-type summation \cite{Pit} for two thin planar surface sheets with a pair interaction of the KS form scaling as ${\cal V}(R) \sim R^{-2}$, gives the interaction pressure as \cite{KS1}
\begin{equation}
p = \frac{F(R)}{S} = - \frac{\partial}{\partial D} \int_D^{\infty} 2\pi R ~dR ~{ {\cal V}(R)} \sim D^{-1}.
\label{fxndhqulch}
￼￼\end{equation}
The two forms of the disjoining pressure, Eqs. \ref{fngewh} and \ref{fxndhqulch}, are thus identical, meaning that the KS interaction is nothing but a monopolar fluctuation interaction. This is clear from the fact that the separation dependence of the fluctuation interaction free energy between two surfaces is slower then in the case of standard vdW interactions that stem from dipolar fluctuations between either two semi-infinite media or two thin layers, scaling respectively as \cite{Pit}
\begin{equation}
p = \frac{F(R)}{S} = - \frac{A(D)}{12 \pi D^2} \quad {\rm and/or} \quad - \frac{2 A(D) a^2}{\pi D^5},
\end{equation}
respectively. 

The KS fluctuation forces thus originate in monopolar fluctuations and follow a different scaling either between point particles, $R^{-2}$, or between fluctuating surface layers, $ D^{-1}$, then in the case of dipolar fluctuations. They arise directly from surface capacitance that is non-zero only for a surface free energy that is non-linear, i.e. at least quadratic, w.r.t. the local electrostatic potential.

\begin{figure*}[!t]
\subfloat[]{\includegraphics[width=0.5\textwidth]{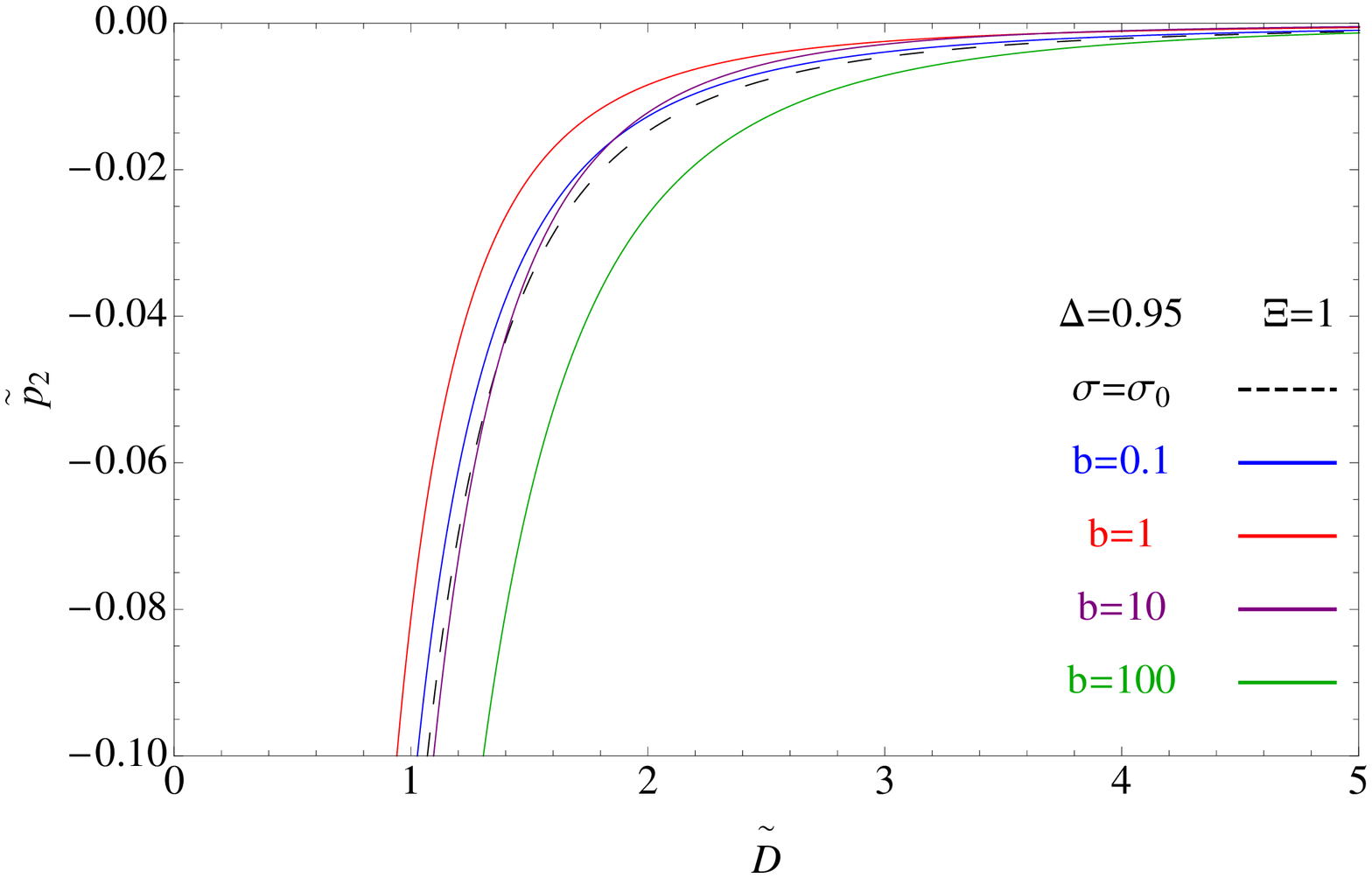}}\subfloat[]{\includegraphics[width=0.49\textwidth]{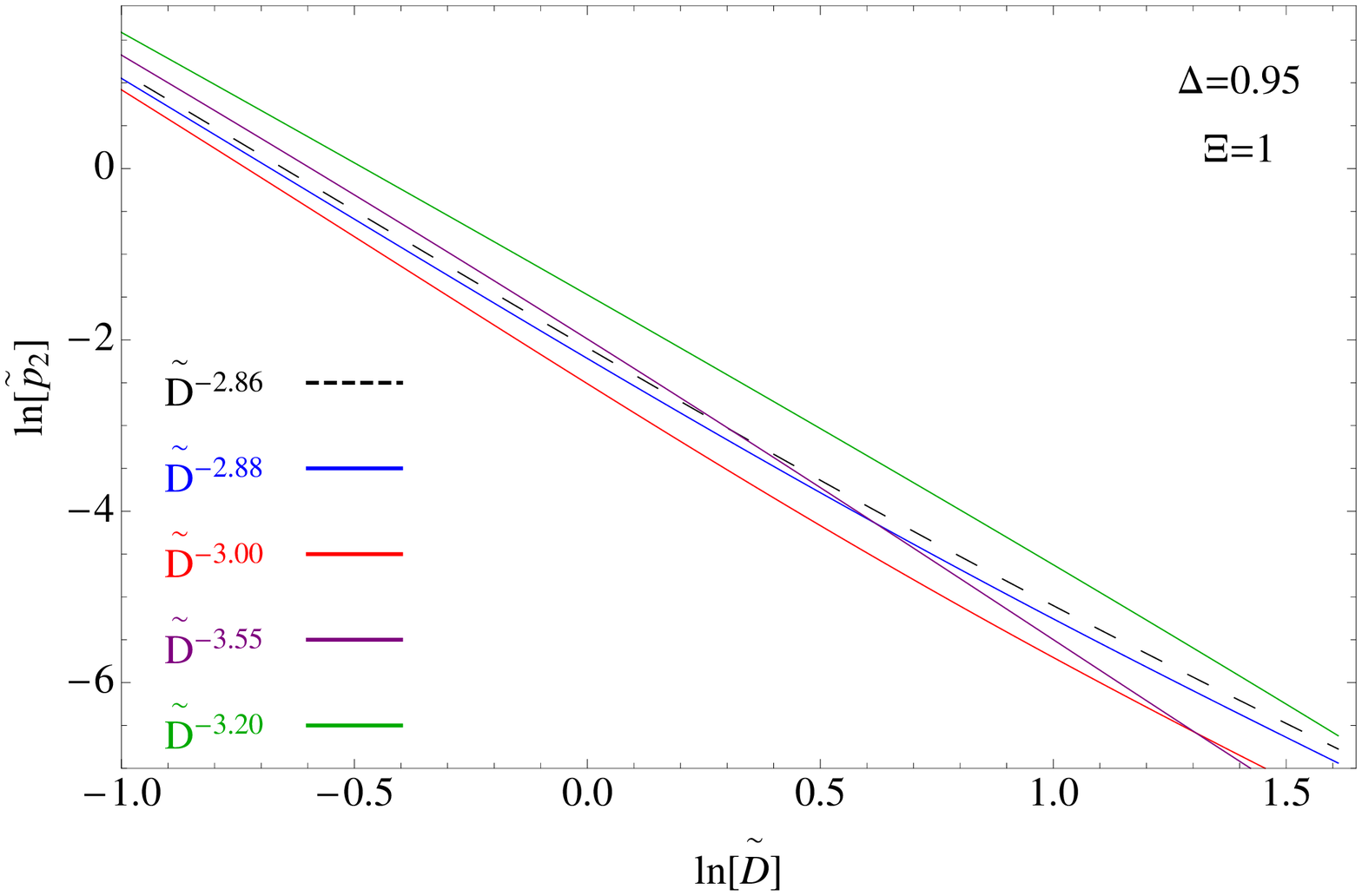}}
\caption{(a) Rescaled fluctuation  disjoining pressure as a function of rescaled surface separation is plotted for different values of parameter $\ b$ with a fixed dielectric jump $\ \Delta =0.95$, and coupling parameter $\ \Xi =1$. (b) Rescaled fluctuation disjoining pressure from (a) plotted in a log-log plot to show the effective scaling of the  disjoining pressure with the intersurface separation. The scaling exponent is typically comparable with the case of the counterion-only Coulomb fluid between two surfaces with fixed charges, which is $-3$, but its exact value depends on $b$.}
\label{fig:fig2}
\end{figure*}

\section{Numerical results}\label{subsec:nmr}

It is convenient to introduce dimensionless quantities by using the Gouy-Chapman length scale $\ \mu $ and $\ \sigma _0^2/2\epsilon \epsilon_{0}$ as the  disjoining pressure scale. Hence, the length scale (${\bf{r}}, D$), the free energy ($\ F$), the  disjoining pressure ($\ p$) and the surface capacitance ($\ {\cal{C}}$) can all be rescaled into dimensionless variables $\ \tilde{\bf{r}}={\bf{r}}/\mu, \tilde D = D/ \mu $, $\  \tilde{F}=F/\left(\frac{\sigma _0^2}{2\epsilon \epsilon_{0}}\right)\mu ^3 $, $\ \tilde{p}=p/\left(\frac{\sigma _0^2}{2\epsilon \epsilon_{0}}\right)$ and $\ {\tilde{\cal{C}}}=\mu {\cal{C}}$ respectively.  We also introduce the dielectric mismatch with $\ \Delta=({\epsilon -\epsilon '})/({\epsilon+\epsilon '})$. With these definitions, the mean-field free energy becomes
\begin{equation}
\frac{{\tilde{\cal F}}_0(\tilde{D})}{\tilde{S}}=\tilde{\alpha }^2\tilde{D}+2\ln{[1+\tilde{\alpha }^2]},
\end{equation}
where $\ \tilde{\alpha }=\mu \alpha $ is the solution of the boundary condition 
\begin{equation}
(1+b)\tilde{\alpha }\tan{(\tilde{\alpha }\tilde{D}/2)}+b\tilde{\alpha }\tan^3{(\tilde{\alpha }\tilde{D}/2)}=1\label{eq:as3}
\end{equation}
The rescaled surface capacitance in terms of $\tilde{\alpha }$ is then equal to
\begin{equation}
\tilde{\cal{C}}_{S_1,S_2}=2\beta \epsilon \epsilon _0 b\frac{1+\tan^2{[\tilde{\alpha }\tilde{D}/2]}}{(1+b+b\tan^2{[\tilde{\alpha }\tilde{D}/2]})^2}
\end{equation}
which goes to zero for large values of $\ b$, $\lim_{b \longrightarrow \infty}\tilde{\cal{C}}_{S_1,S_2} \to 0$ as well as for vanishing $b$, $\lim_{b \longrightarrow 0}\tilde{\cal{C}}_{S_1,S_2} \to 0$.
We also invoke a coupling parameter $\ \Xi $, analogous to the one introduced by Netz and Moreira \cite{couplingpar}, given as
 \begin{equation}
 \Xi =\frac{e_0^3\sigma _0}{8\pi (\epsilon \epsilon _0k_BT)^2}
 \end{equation}
for monovalent counterions.  For a counterion only-system with fixed surface charge the magnitude of the coupling parameter defines a weak- and a strong-coupling regime \cite{RudiandCo}. In our case the existence of the surface free energy introduces also other length scales that preclude a direct introduction of a unique electrostatic coupling parameter and it is thus in general not possible to establish the presence of the weak and the strong coupling limits {\sl strictu senso} as exact limits of the partition function. 

\begin{figure*}
\subfloat[]{\includegraphics[width=0.5\textwidth]{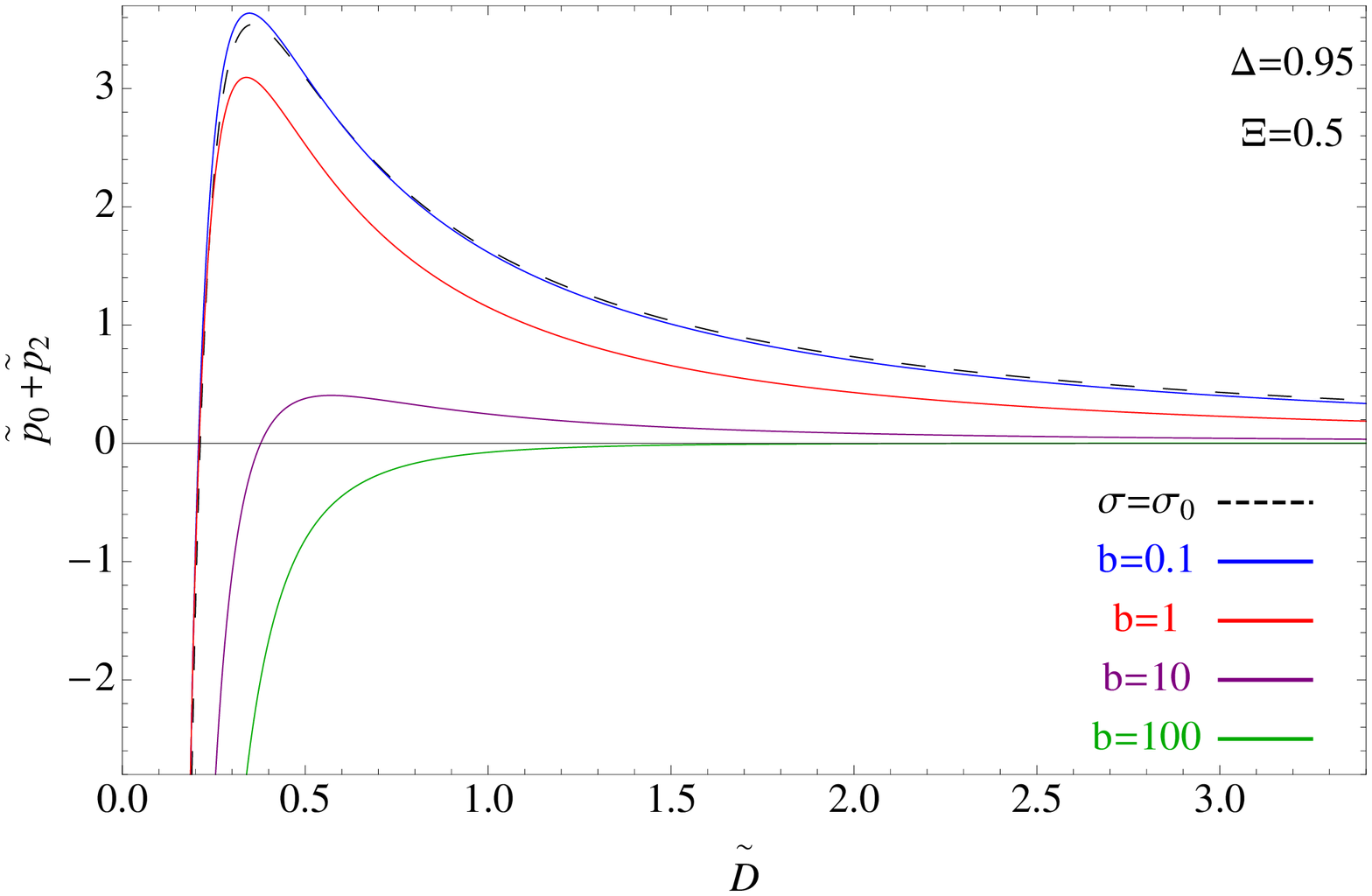}}\subfloat[]{\includegraphics[width=0.5\textwidth]{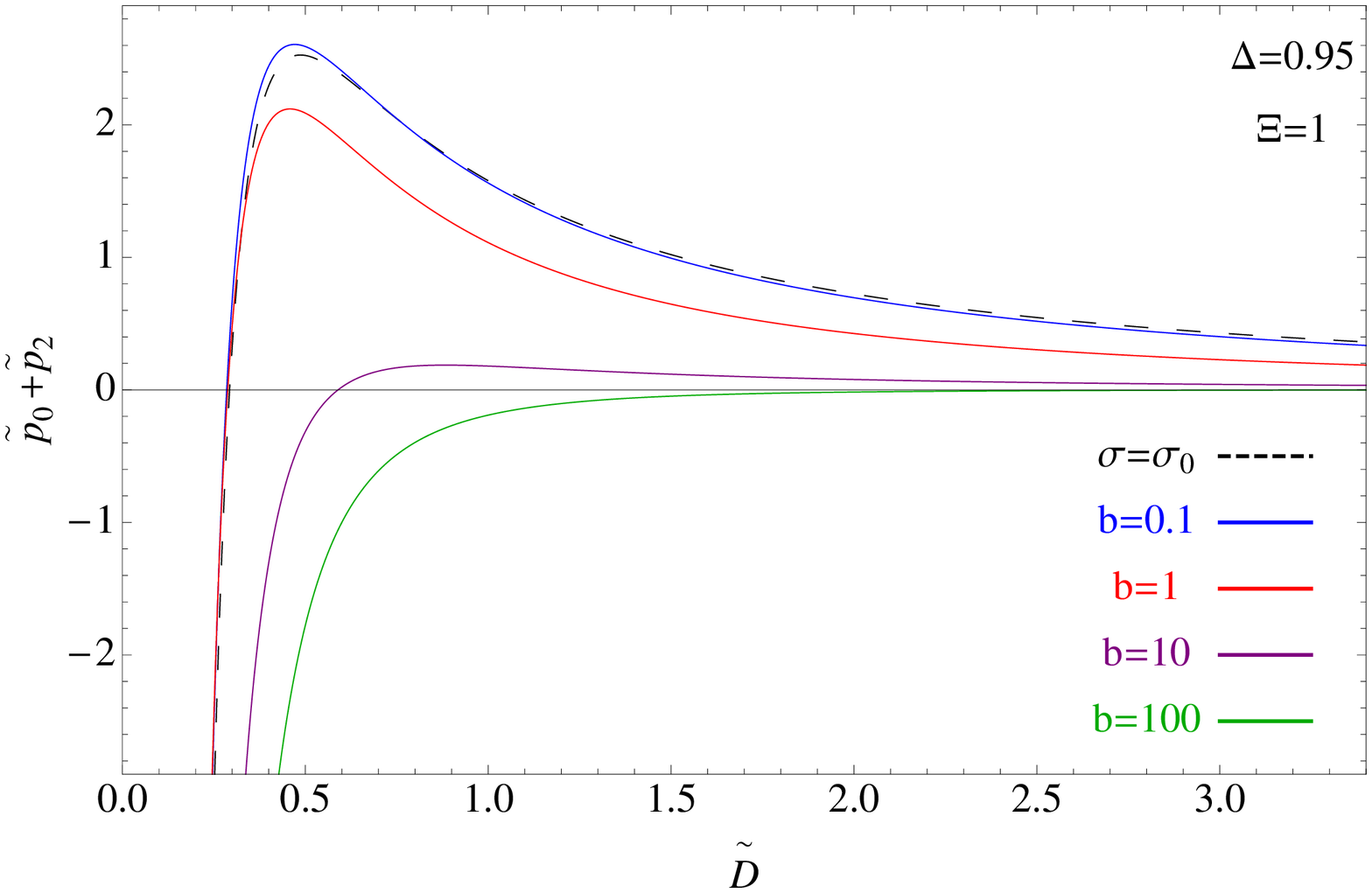}}
\caption{Rescaled total  disjoining pressure as a function of the rescaled surface separation plotted for different values of the parameter $\ b$, fixed dielectric jump $\ \Delta =0.95$ and for the following values of the coupling parameter:  (a) $\ \Xi =0.5$; (b) $\ \Xi =1$.}
\label{fig:fig32}
\end{figure*}
\begin{figure*}[!t]
\subfloat[]{\includegraphics[width=0.51\textwidth]{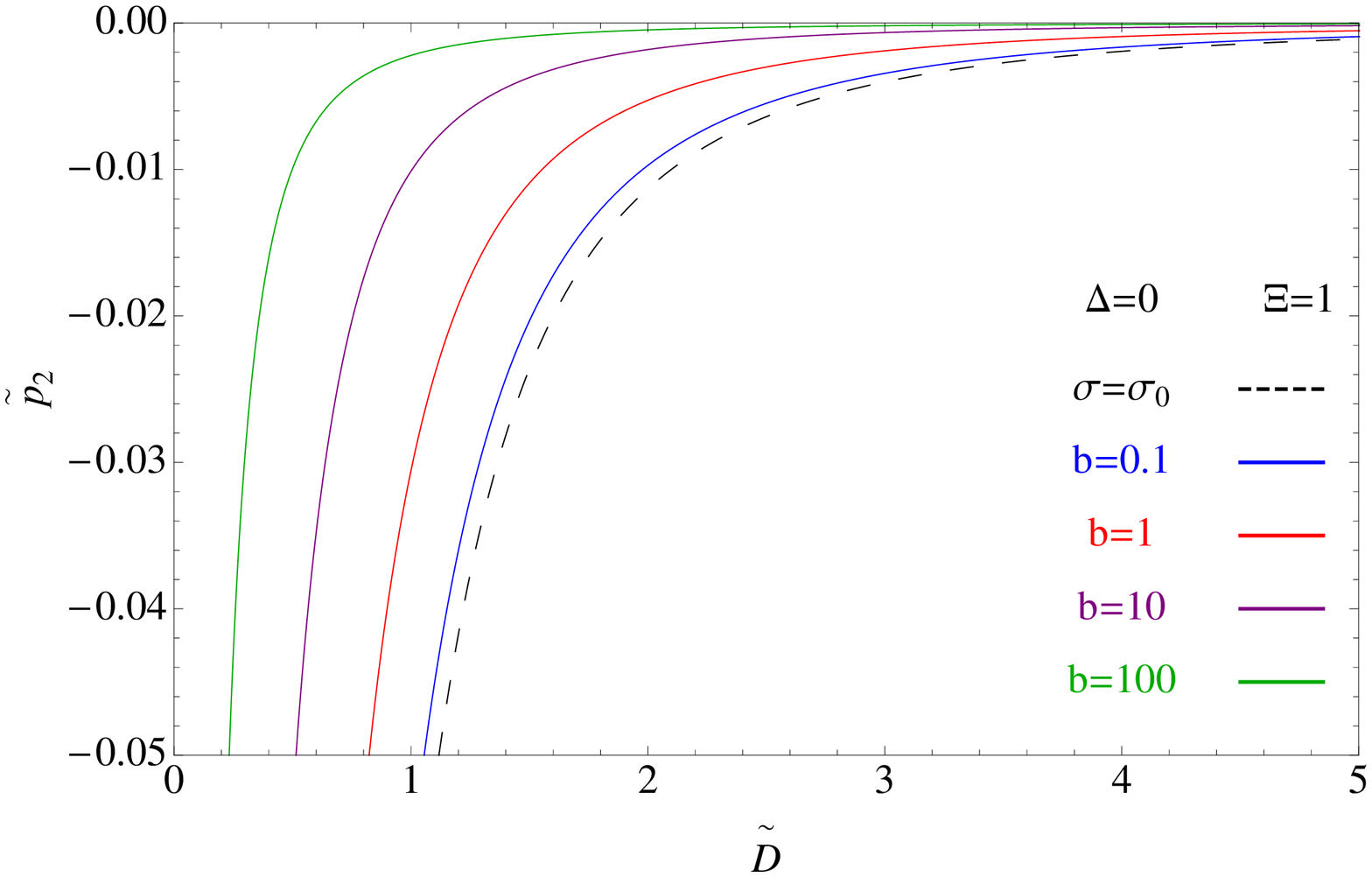}}\subfloat[]{\includegraphics[width=0.5\textwidth]{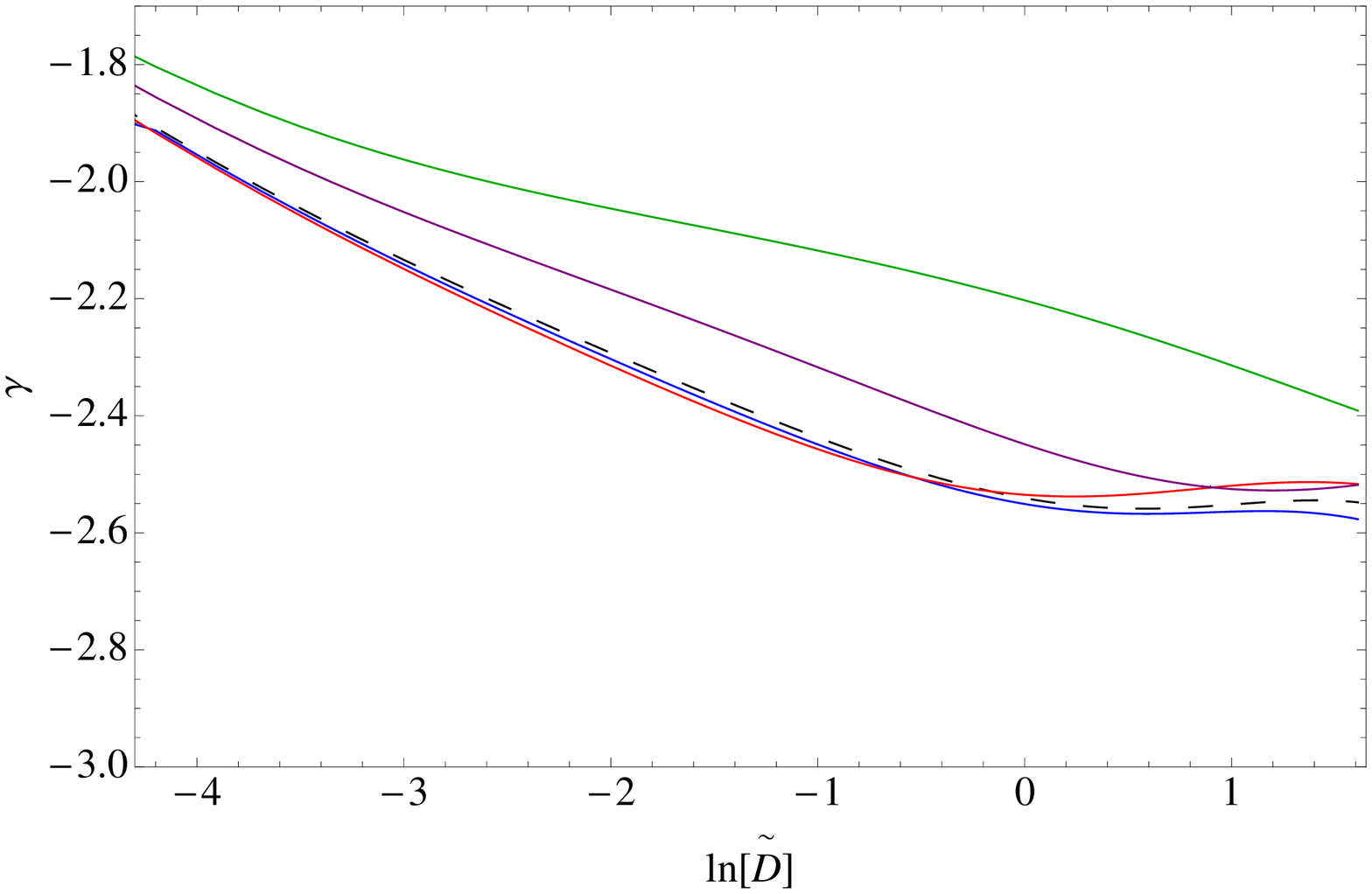}}
\caption{(a) Rescaled fluctuation  disjoining pressure as a function of rescaled surface separation is plotted for different values of the parameter $\ b$ but without any dielectric jump, $\ \Delta =0$, and $\ \Xi =1$. (b) The scaling exponent $\ \gamma $ for the effective scaling of the  disjoining pressure with the intersurface separation is defined as $\ \tilde{p}_2\sim \tilde{D}^{\gamma }$.  For small separations it approaches $-1$ asymptotically, whereas for large separations it tends to a value close but not equal to $-3$.}
\label{fig:fig3}
\end{figure*}

While the weak coupling limit can therefore not be derived as an exact limit, the saddle-point can be defined for any field-action. As explained in detail in Ref. \cite{RudiandCo} we thus use the saddle-point solution as the {\sl proxy} for the weak coupling limit and evaluate the contribution of the fluctuations around the saddle-point to the free energy. 

The surface interaction part of the Gaussian fluctuating free energy from Eq. \ref{eq:tr3dksl}, in a dimensionless form is then given as:
\begin{widetext}
\begin{eqnarray}\label{eq:tr3dkslpoqwv1}
&&\frac{{\tilde{\cal F}}_2(\tilde{D})}{\tilde{S}}={\textstyle{\frac{1}{2}}}\Xi \int_0^\infty\tilde{Q}d\tilde{Q}~ \ln{\Big[ 1 - e^{- 2\tilde{Q}\tilde{D} }}\times\nonumber\\
&&\!\!\!\!\!\times\!\!\!\left(\!\!\frac{2(1+\Delta )b(1+\tan^2{[\frac{\tilde{\alpha }\tilde{D}}{2}]})(\tilde{\alpha }\tan{[\frac{\tilde{\alpha }\tilde{D}}{2}]}-\tilde{Q})\!\!-\!\!(1+b+b\tan^2{(\frac{\tilde{\alpha }\tilde{D}}{2})})^2[2\Delta \tilde{Q}(\tilde{\alpha }\tan{[\frac{\tilde{\alpha }\tilde{D}}{2}]}-\tilde{Q})\!\!-\!\!(1+\Delta )(\tilde{\alpha }^2+\tilde{\alpha }^2\tan^2{[\frac{\tilde{\alpha }\tilde{D}}{2}]})]}{2(1+\Delta )b(1+\tan^2{[\frac{\tilde{\alpha }\tilde{D}}{2}]})(\tilde{\alpha }\tan{[\frac{\tilde{\alpha }\tilde{D}}{2}]}+\tilde{Q})\!\!+\!\!(1+b+b\tan^2{(\frac{\tilde{\alpha }\tilde{D}}{2})})^2[2\tilde{Q}(\tilde{\alpha }\tan{[\frac{\tilde{\alpha }\tilde{D}}{2}]}+\tilde{Q})\!\!+\!\!(1+\Delta )(\tilde{\alpha }^2+\tilde{\alpha }^{\!\!\!2}\tan^2{[\frac{\tilde{\alpha }\tilde{D}}{2}]})]}\!\!\right)^{\!\!2}\!\!\!\!\Big]\nonumber\\
~
\end{eqnarray}
\end{widetext}

We first investigate the surface separation dependence of the interaction free energy and the disjoining  pressure between the surfaces pertaining to that dependence. The mean-field rescaled pressure is shown in Fig. \ref{fig:fig1} (a), as a function of the surface dissociation energy $\ln{b}= \beta \mu_S$ in a lin-lin and log-log plots. Clearly, the higher the energy penalty for charge dissociation at the surface, $b$, the lower is the interaction pressure between the two surfaces until for large enough energy penalty the interaction remains close to zero for all intersurface separations. The scaling of the mean-field  disjoining pressure with the separation  is shown in Fig. \ref{fig:fig1} (b). For constant surface charge $\sigma = \sigma_0$, i.e., corresponding formally to $b = 0$, the asymptotic forms of the mean-field interaction pressure are $\lim_{D\longrightarrow \infty} \tilde p_0(D) \sim {\tilde D}^{-2}$ and $\lim_{D\longrightarrow 0} \tilde p_0(D) \sim {\tilde D}^{-1}$, see Ref. \cite{Andelman2}. This is in fact also what we observe in the case of charge regulation, with the proviso that  the regime of validity of the two limits depends additionally on the value of $b$; the smaller its value the more extended is the region of ${\tilde D}^{-1}$ scaling.

Because the surface capacitance depends on the mean-field solution, the fluctuation correction to the free energy and the corresponding  disjoining pressure also depend on the surface dissociation energy, as can be discerned from Fig. \ref{fig:fig2} (a). This is very different from the standard vdW interactions that do not depend on the mean-field solution, at least in the standard DLVO formulation \cite{Pit}.  The scaling of the fluctuation part of the interaction pressure, Fig. \ref{fig:fig2} (b), shows a robust value of the scaling exponent close to $-3$, close to its value for the case of a counterion-only Coulomb fluid between two surfaces with fixed charges, where the fluctuation disjoining pressure scales exactly as $\sim \log{D} \times D^{-3}$, see Ref. \cite{RudiandCo} for details. The exact value of the scaling exponent in the charge-regulated case, however, depends on the value of the surface interaction parameter $b$. Since the dielectric mismatch in this case is not zero, the monopolar and vdW dipolar fluctuation interactions, stemming from the surface capacitance and the dielectric mismatch respectively,  are always mixed together and can not be disentangled in the separation dependence of the fluctuation pressure.

Adding the mean-field and the fluctuation contribution together, Fig. \ref{fig:fig32}, we note that for large values of the surface dissociation energy, the fluctuation contribution becomes dominant, a simple consequence of the fact that the mean-field vanishes while the fluctuation part remains finite. While in general the fluctuation part is always subdominant to the mean-field solution, in this case the matters are a bit more complicated as the charge regulation can wipe out the mean-field entirely but not the fluctuation part. The fluctuation  disjoining pressure for a vanishing mean-field again depends crucially on the presence of the dielectric mismatch at the bounding surfaces and does not necessarily coincide with the standard vdW interaction. In fact for the case of complete dielectric homogeneity, $\Delta = 0$ see Fig. \ref{fig:fig3}, the interaction pressure scaling exponent is in general smaller then for $\Delta \neq 0$. Asymptotically for small separations in fact it approaches one, just as for the KS interaction. For larger separations it tends to a larger value but does not approach $-3$ as the fluctuations it corresponds to, being due to the presence of counter ions between the surfaces,  are never purely dipolar. 

Finally, in order to get an idea about the strength of the attractive interaction we compare the fluctuation disjoining pressure $\ p_2$ with the pure van der Waals pressure given as $\ p_{vdW}=-H(\Delta)/12\pi D^3$, see Ref. \cite{Pit}, where $\ H(\Delta)$ is a Hamaker coefficient, which for illustration purposes we chose to be 4.3 zJ \cite{Rudinon}. We choose a large dielectric inhomogeneity ($\ \Delta=0.95$), and a separation between the surfaces of 1 nm ($\ D=1$ nm), bearing maximal surface charge $\ \sigma_0=0.5e_0/nm^2$. With the given set of parameters, we calculate the fluctuation disjoining pressure $\ p_{b=0}$ corresponding to a maximal charge at the surfaces, and the fluctuating disjoining pressure $\ p_{b=100}$, corresponding to the case of electroneutral surfaces. One finds that for this specific choice of parameters the fluctuating pressure is comparable to the vdW disjoining pressure: $\ p_{vdW}=-1.1 atm$ while $\ p_{b=100}=-1.3 atm$ and $\ p_{b=0}=-0.8atm$.

\section{Conclusion}\label{sec:con}

In this paper we derived a theory describing electrostatic interactions between macromolecular surfaces bearing dissociable charge groups immersed in an aqueous solution of dissociated counterions. Introducing a surface free energy corresponding to a simple model of charge regulation, and formulating it in a field-theoretic language, we derived the mean-field solution which is related to the Ninham-Parsegian charge regulation theory and also obtained an exact solution for the second order fluctuations around the mean field. The fluctuation contribution to the total free energy is related to vdW interactions but is fundamentally modified by the presence of dissociable charges on the bounding surfaces as well as the counter ions dissolved in the space between them. 

While for the model discussed, containing an additional surface term usually not present in Coulomb fluids with fixed charges on interacting surfaces, a weak-coupling approximation can not be consistently defined, we proceed from the observation that the saddle-point and the fluctuations around the saddle point can be defined for any field action \cite{RudiandCo}. The range of validity of this approximation should eventually be ascertained once compared with detailed simulations of the same microscopic model. 

What our methodology also clearly identifies is the monopolar nature of the fluctuation interactions between charge-regulated surfaces that singles them out from the dipolar fluctuation interactions as is the case for vdW fluctuation interactions. This sets the two types of interactions fundamentally apart as the range and scaling characteristics of the two are vastly different. It also emerges quite straightforwardly that the two types of fluctuation interactions are not additive but are fundamentally intertwined and can only be decoupled in extreme limiting cases of either no dielectric discontinuity or in the case of no surface capacitance. More specific predictions regarding the role of monopolar fluctuation interactions between dissociable charge groups corresponding to deprotonated and protonated molecular groups, as is the case for proteins, will be forthcoming once the model considered is generalized to include the intervening salt solution at a set value of the solution pH.
  
Suffice it to say at this point that in an appropriate limit our theory is related to the KS interactions known to be relevant in the protein context. More importantly though, it allows to consistently generalize the theory of KS interactions, or indeed any electrostatic interaction that presumes charge regulation, in such a way that one can use advanced concepts and methods of the Coulomb fluid theory to solve it approximately. In this way we pave the way to new developments in the theory of KS and related interactions that would not be conceivable within their original theoretical framework \cite{KS1,KS2}. The field-theoretic framework in fact allows to formulate a single-particle partition function which can be used as a proxy for the strong-coupling approximation, also not consistently defineable in the case where the field action contains additional surface terms, as in the model introduced here \cite{RudiandCo}. We are currently working to extend the present formulation to the case of symmetric as well as asymmetric ionic mixtures containing monovalent and polyvalent ions.

\section{Acknowledgments}

N. A. is grateful to An\v{z}e Lo\v{s}dorfer Bo\v{z}i\v{c} for constructive advice regarding this work. N. A. acknowledges the financial support by the Slovenian Research Agency under the young researcher grant. R.P. acknowledges the financial support by the Slovenian Research Agency under the grant P1-0055. R.P. would like to thank the hospitality of Prof. R.R. Netz during his stay at the Freie Universit\" at and the Technische Universit\" at in Berlin as a visiting professor, where parts of this work were performed.

\begin{appendix}
\section{Exact evaluation of the path integral}\label{sec:ap1}

The path integral in Eq. \ref{eq:pathkl} can be written in the form \cite{KL}:
\begin{eqnarray}\label{eq:prop}
&& {\cal G}_p\Big({\delta \phi}(Q,z_1), {\delta \phi}(Q,z_2)\Big) = \nonumber\\
&& \sqrt{\frac{ 1}{2\pi }} \exp{\Big[-\frac{1}{2}\int_{-d}^{d}dz\int_0^1d\mu \mathcal{R}(z,z,\mu )\Big]}\times\nonumber\\
&&\times\exp{\Big[ - \frac{\beta \epsilon\epsilon_0 }{2 }\Big(\delta \phi (Q, z_2)f'(z_2) - {\delta \phi }(Q,z_1)f'(z_1)\Big)\Big]},\nonumber\\
~
\end{eqnarray}
where $\ f(z)$ is a solution of the equation of motion given as:
\begin{equation}
\ddot{f}-\mu \Big(Q^2 + \frac{2\alpha ^2}{\cos^2{(\alpha z)}}\Big)f=0,\label{eq:ddg}
\end{equation}
where $\ f=f(z;\mu )$. The Green' s function equation is:
\begin{equation}
\frac{d^2}{dz^2}Q(z,z'|\mu )-\mu \Big(Q^2 + \frac{2\alpha ^2}{\cos^2{(\alpha z)}}\Big)Q(z,z'|\mu )=-\delta (z-z'),
\end{equation}
with $\ Q(-d,z'|\mu )=Q(d,z'|\mu)=0$. The resolvent $\ \mathcal{R}(z,z'|\mu )$ obeys the equation:
\begin{eqnarray}
&&\frac{d^2}{dz^2}\mathcal{R}(z,z'|\mu )-\mu \Big(Q^2 + \frac{2\alpha ^2}{\cos^2{(\alpha z)}}\Big)\mathcal{R}(z,z'|\mu )=\nonumber\\
&&=\delta (z-z')\Big(Q^2 + \frac{2\alpha ^2}{\cos^2{(\alpha z)}}\Big),
\end{eqnarray}
with $\ \mathcal{R}(-d,z'|\mu )=\mathcal{R}(d,z'|\mu)=0$. We can see that the resolvent satisfies $\ \mathcal{R}(z,z'|\mu )=-\Big(Q^2 + \frac{2\alpha ^2}{\cos^2{(\alpha z')}}\Big)Q(z,z'|\mu )$. The Green' s function $\ Q(z,z'|\mu )$ has the form:
\begin{eqnarray}
Q(z,z'|\mu )=&&\begin{cases}
g(z,\mu )h(z',\mu )/\Delta (\mu ),  		z<z'
\end{cases}\nonumber\\&&
\begin{cases}
g(z',\mu )h(z,\mu )/\Delta (\mu ),  		 z>z'
\end{cases}\nonumber\\
~
\end{eqnarray}
where $\ g(z,\mu )$ and $\ h(z,\mu )$ are two linearly independent solutions of Eq. \ref{eq:ddg} satisfying the conditions:
\begin{equation}
g(-d;\mu )=h(d;\mu )=0\label{eq:dgbc}
\end{equation}
and
\begin{equation}
\Delta (\mu )=\dot{g}(-d,\mu )h(-d,\mu )=-g(d,\mu )\dot{h}(d,\mu )
\end{equation}
The integration of the resolvent operator yields:
\begin{eqnarray}
&&\int_{-d}^d\mathcal{R}(z,z|\mu ) dz =-\int_{-d}^d\Big(Q^2 + \frac{2\alpha ^2}{\cos^2{(\alpha z)}}\Big)Q(z,z|\mu )dz=\nonumber\\
&&[-1/\Delta (\mu )]\int_{-d}^d\Big(Q^2 + \frac{2\alpha ^2}{\cos^2{(\alpha z)}}\Big)g(z,\mu )h(z,\mu )dz.\nonumber\\
~
\end{eqnarray}
Consider now the equation satisfied by g:
\begin{equation}
\ddot{g}-\mu \Big(Q^2 + \frac{2\alpha ^2}{\cos^2{(\alpha z)}}\Big)g=0
\end{equation}
and differentiating it with respect to $\ \mu $, we have:
\begin{equation}
-\Big(Q^2 + \frac{2\alpha ^2}{\cos^2{(\alpha z)}}\Big)g=\ddot{g}_{\mu }+\mu \Big(Q^2 + \frac{2\alpha ^2}{\cos^2{(\alpha z)}}\Big)g_{\mu }.\nonumber\\
~
\end{equation}
Inserting this into the resolvent integral and integrating by parts, one can get:
\begin{eqnarray}
&&\int_{-d}^d\mathcal{R}(z,z|\mu )dz=[-1/\Delta(\mu )]\times\nonumber\\
&&\times[\dot{g}_{\mu }(-d,\mu )h(-d,\mu )-\dot{h}(d,\mu )g_{\mu }(d,\mu )]=\nonumber\\
&&=-g_{\mu }(d,\mu )/g(d,\mu )-\dot{g}_{\mu }(-d,\mu )/\dot{g}(-d,\mu ),\nonumber\\
~
\end{eqnarray}
from which it follows that:
\begin{eqnarray}
&&\int_0^1d\mu \int_{-d}^d dz\mathcal{R}(z,z|\mu )=\ln{[g(d,\mu )/\dot{g}(-d,\mu )]}\vert_0^1=\nonumber\\
&&\ln{[(g(d,1)/\dot{g}(-d,1))(\dot{g}(-d,0)/g(d,0))]}.\nonumber\\
~
\end{eqnarray}
As $\ g(d,0)/\dot{g}(-d,0)=2d=D$, we have:
\begin{equation}
\exp{[-\frac{1}{2}\int_0^1d\mu \int_{-d}^d dz\mathcal{R}(z,z|\mu )]}=[D\dot{g}(-d,1)/g(d,1)]^{\frac{1}{2}}.
\end{equation}
Now, the solution of the equation of motion is given as a linear combination of the solutions $\ g $ and $\ h$ as:
\begin{equation}
f(z,1)=\delta \phi _2g(z,1)/g(d,1)+\delta \phi _1h(z,1)/h(-d,1),\nonumber\\
~
\end{equation}
so the exponent in the propagator Eq. \ref{eq:prop} becomes:
\begin{eqnarray}
&&\exp{\Big[ - \frac{\beta \epsilon\epsilon_0 }{2 }\Big(\delta \phi (Q, z_2)f'(z_2) - {\delta \phi }(Q,z_1)f'(z_1)\Big)\Big]}=\nonumber\\
&&\exp{\Big[ - \frac{\beta \epsilon\epsilon_0 }{2 }\Big(\delta \phi ^2(Q, d)\frac{\dot{g}(d,1)}{g(d,1)} -}\nonumber\\
&&{-2\delta \phi (Q,d)\delta \phi (Q,-d)\frac{\dot{g}(-d,1)}{g(d,1)} -{\delta \phi ^2}(Q,-d)\frac{\dot{h}(-d,1)}{h(-d,1)}\Big)\Big]}.\nonumber\\
~
\end{eqnarray}
Finally the propagator can be written as: 
\begin{eqnarray}\label{eq:propagator}
&& {\cal G}_p\Big({\delta \phi}(Q,-d), {\delta \phi}(Q,d)\Big) = \nonumber\\
&& \sqrt{\frac{D\dot{g}(-d,1)}{2\pi g(d,1)}}\exp{\Big[ - \frac{\beta \epsilon\epsilon_0 }{2 }\Big(\delta \phi ^2(Q, d)\frac{\dot{g}(d,1)}{g(d,1)} -}\nonumber\\
&&{-2\delta \phi (Q,d)\delta \phi (Q,-d)\frac{\dot{g}(-d,1)}{g(d,1)} -{\delta \phi ^2}(Q,-d)\frac{\dot{h}(-d,1)}{h(-d,1)}\Big)\Big]}.\nonumber\\
~
\end{eqnarray}
Solutions $\ g(z,1)$ and $\ h(z,1)$, which satisfy equation Eq. \ref{eq:ddg} when $\ \mu =1$ and boundary conditions Eq. \ref{eq:dgbc}, are given as:
\begin{eqnarray}
g(z)&=&\frac{\sinh{[Q(d+z)]}(Q^2\cot{[\alpha d]}+\alpha ^2\tan{[\alpha z]})}{\alpha (Q^2+\alpha ^2)}+\nonumber\\
&+&\frac{\alpha Q\cosh{[Q(d+z)]}(1+\cot{[\alpha d]}\tan{[\alpha z]})}{\alpha (Q^2+\alpha ^2)};\nonumber\\
h(z)&=&\frac{\sinh{[Q(-d+z)]}(Q^2\cot{[\alpha d]}-\alpha ^2\tan{[\alpha z]})}{\alpha (Q^2+\alpha ^2)}+\nonumber\\
&+&\frac{\alpha Q\cosh{[Q(-d+z)]}(-1+\cot{[\alpha d]}\tan{[\alpha z]})}{\alpha (Q^2+\alpha ^2)}.\nonumber\\
~
\end{eqnarray}
After inserting  these solutions back into the equation Eq. \ref{eq:propagator}, one obtains the final result in the explicit form: 
\begin{widetext}
\begin{eqnarray}\nonumber
&& {\cal G}_Q\Big({\delta \phi}(Q,-\frac{D}{2}), {\delta \phi}(Q,\frac{D}{2})\Big) = \sqrt{\frac{A}{2\pi B}}\times \exp{\Big[ - \frac{\beta \epsilon\epsilon_0 }{2B}} \Big([\delta \phi ^2(Q,-\frac{D}{2})+\delta \phi ^2(Q,\frac{D}{2})]C-2\delta \phi (Q,-\frac{D}{2})\delta \phi (Q,\frac{D}{2})A\Big)\Big], \label{eq:gfprop}
\end{eqnarray}
\end{widetext}
where $z_2 = D/2$, $z_1 = -D/2$ while $\ A$, $\ B$ and $\ C$ are defined as
\begin{widetext}
\begin{eqnarray}
A&=&Q(\alpha ^2+Q^2)\cot^2{(\alpha D/2)};\nonumber\\ 
B&=& 2\alpha Q\cosh{(DQ)}\cot{(\alpha D/2)}+(\alpha ^2+Q^2\cot^2{(\alpha D/2)})\sinh{(DQ)};\nonumber\\
C&=& Q\cosh{(DQ)}(2\alpha ^2+(\alpha ^2+Q^2)\cot^2{(\alpha D/2)})+ 2\alpha (\alpha ^2+Q^2+Q^2\cos{(\alpha D)})\csc{(\alpha D)}\sinh{(DQ)}.
\end{eqnarray}
\end{widetext}
\end{appendix}

{REFFERENCES:}

\end{document}